\documentclass[
aps,     
prl,                    
showpacs,               
superscriptaddress,     
nofootinbib,            
twocolumn,               %
floatfix, nobalancelastpage
]{revtex4}

\usepackage[T1]{fontenc}
\usepackage[latin9]{inputenc}
\usepackage{amsmath}
\usepackage{graphicx}
\usepackage{tikz-feynman}
\usepackage{simplewick}
\usepackage{slashed}
\usepackage{subfigure}

\begin{document}

\preprint{\vbox{\hbox{WSU-HEP-1805}}}

\title{\boldmath Invisible widths of heavy mesons}

\author{Bhubanjyoti Bhattacharya}
\affiliation{Department of Natural Sciences\\
        Lawrence Technological University, Southfield, MI 48075, USA}

\author{Cody M. Grant}
\affiliation{Department of Physics and Astronomy\\
        Wayne State University, Detroit, MI 48201, USA}

\author{Alexey A.\ Petrov}
\affiliation{Department of Physics and Astronomy\\
        Wayne State University, Detroit, MI 48201, USA}


\begin{abstract}
\noindent
We revisit calculations of invisible widths of heavy mesons in the standard model, which serve as benchmarks for the
studies of production of light, long-lived neutral particles in heavy meson decays.
We challenge the common assumption that in the standard model these widths are dominated by meson decays into a
two-neutrino final state and prove that they are dominated by decays into four-neutrino final states. We show that
current estimates of the invisible widths of heavy mesons in the standard model underestimate the effect by orders of
magnitude. We examine currently available experimental data on invisible widths and place constraints on the
properties of dark photons. We also comment on the invisible widths of the kaons.
\end{abstract}

\maketitle


Experimental studies of light, $m \sim {\cal O}(0.1-10^3)$ MeV, weakly interacting long lived particles (WILLPs) have received
considerable attention recently, in part due to development of new models of dark matter (DM). These particles could help resolve
several problems in physics of dark matter, both by being DM states and/or serving as mediators between visible and dark sectors
of our Universe \cite{Craig:2015pha}. As such, extensive experimental programs of searches for light
WILLPs \cite{Meade:2009mu,ATLAS:2012av,Aad:2015uaa,Aaij:2017mic} have been put forward at several experimental
centers around the world. For recent constraints on candidates for the light particles such as axion-like states or dark photons
see \cite{Essig:2013lka,Patrignani:2016xqp}.

If these WILLP states exist, they could also be produced in the decays of mesons, such as $B$, $D$, or even $K$. One of the
tantalizing experimental signatures of such transitions includes ``invisible'' decays of heavy meson states \cite{Badin:2010uh},
as light WILLPs do not interact with the detectors. Currently operating experiments Belle II and BESIII at $e^+e^-$ machines in
Japan and China, as well as experiments at future flavor factories, are the ideal places for experimental studies of such decays.
This is because flavor factories operate at the $\Upsilon (4S)$ ($b\bar b$) or $\psi(3770)$ ($c\bar c$) resonances that decay into
a correlated state of $B_d^0 (D^0)$ meson pairs. Thus, ``invisible'' $B^0_d(D^0)$ decays into WILLPs can be tagged with
non-leptonic decays of $\bar B_d^0 (\bar D^0)$ decays ``on the other side.''

Current experimental constraints on those decays come from the analyses done at BaBar and Belle collaborations (for $B^0$) and by
Belle collaboration (for $D^0$). No signals have been observed so far, so upper limits on the branching fractions of heavy mesons
decaying to invisible final states ${\cal B}(B_{d}^0 \to \slashed{E}) < 1.3 \times 10^{-4}$ (Belle) \cite{Hsu:2012uh} and ${\cal B}
(B_{d}^0 \to \slashed{E}) < 2.4 \times 10^{-5}$ (BaBar) \cite{Lees:2012wv} for the $b$-flavored mesons and ${\cal B}(D^0 \to
\slashed{E}) < 9.4 \times 10^{-5}$ (Belle) \cite{Lai:2016uvj} for charmed mesons have been set at 90\% confidence level.

If measurements of invisible width of a heavy meson are to be used in placing constraints on new physics models \cite{Badin:2010uh},
standard model (SM) backgrounds to those modes need to be estimated. While different experiments have different experiment-specific
backgrounds for such processes related to ``missing'' particles in their detectors \cite{Hsu:2012uh,Lees:2012wv,Lai:2016uvj} that
can be controlled with various experimental methods, the irreducible SM backgrounds to invisible meson decays have not received
complete attention in the theoretical literature.

The only irreducible SM background that has the same experimental signature is heavy meson decays into the final states containing
only neutrinos. Transitions of a $B^0_q (D^0)$ meson into such final states are described by an effective Lagrangian,
\begin{eqnarray}\label{eqn:FCNC}
{\cal L}_{eff} &=& -\frac{4 G_F}{\sqrt{2}}\frac{\alpha}{2\pi\sin^2\theta_W}
\nonumber \\
&\times&
\sum_{l=e,\mu,\tau} \sum_{k} \lambda_k X^l(x_k)
\left( J_{Qq}^\mu\right)
\left(\overline{\nu}^l_L \gamma_\mu \nu^l_L\right),
 \end{eqnarray}
where $J_{Qq}^\mu=\overline{q}_L \gamma^\mu b_L$ for beauty, and $J_{Qq}^\mu=\overline{u}_L \gamma^\mu c_L$ for charm transitions,
and we consider Dirac neutrinos. The functions $\lambda_k X^l(x_k)$ are combinations of the Cabbibo-Kobayashi-Maskawa (CKM) factors
and Inami-Lim functions. These functions are dominated by the top-quark contribution for $b \to q$ transitions, so
\begin{equation}
\sum_{k} \lambda_k X^l(x_k) = V_{tq}^* V_{tb} X(x_t),
\end{equation}
where $x_t=m_t^2/M_W^2$ and
\begin{equation}\label{BqFunct}
X(x_t) = \frac{x_t}{8}\left[\frac{x_t+2}{x_t-1}+\frac{3(x_t-2)}{(x_t-1)^2}\ln{x_t}\right]
 \end{equation}
Perturbative QCD corrections~\cite{Buchalla:1993bv} would numerically change
Eq.~(\ref{BqFunct}) by at most 10\%, so therefore will be neglected. For charm $c \to u$
transitions we keep the contributions from both internal $b$ and $s$-quarks,
\begin{equation}
\sum_{k} \lambda_k X^l(x_k) = V_{cs}^* V_{us} X^l(x_s)+V_{cb}^* V_{ub} X^l(x_b),
\end{equation}
where $X^l(x_q) = \overline D(x_q,y_l)/2$ with $y_l=m_l^2/m_W^2$ are related to the Inami-Lim
functions~\cite{Inami:1980fz},
\begin{eqnarray}
&& \overline D(x_q,y_l)=
\frac{1}{8}\frac{x_q y_l}{x_q-y_l} \left(\frac{y_l-4}{y_l-1}\right)^2 \ln y_l
\nonumber\\
&&~+\frac{x_q}{8} \left[\frac{x_q}{y_l-x_q}\left(\frac{x_q-4}{x_q-1}\right)^2+1+\frac{3}{(x_q-1)^2}\right]\ln{x_q}\nonumber\\
 && ~+ \frac{x_q}{4}-\frac{3}{8}\left(1+3\frac{1}{y_l-1}\right)\frac{x_q}{x_q-1}
 \end{eqnarray}
Given this, one can easily estimate branching ratios for $B_q(D) \to \nu\overline{\nu}$ decays. One can
immediately notice that the left-handed structure of the Lagrangian results in helicity suppression
of these decays due to the fact that initial state is a spin-0 meson. The branching ratio is
 \begin{equation}\label{eqn:Bto2nu}
{\cal B}(B_{q}\rightarrow \nu\overline{\nu})= \frac{G_F^2\alpha^2f_{B_q}^2 M_{B_q}^3} {16 \pi^3 \sin^4
\theta_W\Gamma_{B_q}} |V_{tb}^{}V_{tq}^*|^2X(x_t)^2 x_\nu^2 ,
 \end{equation}
where $x_\nu = m_\nu/M_{B_q}$ and $\Gamma_{B_q}=1/\tau_{B_{q}}$ is the total width of the $B_q$ meson.
We also summed over all possible neutrino states, i.e. $\nu\bar\nu = \sum_{i=1}^3 \nu_i\bar\nu_i$.

As can be seen from Eq.~(\ref{eqn:Bto2nu}), the branching ratio is exactly zero in the minimal standard model with
massless neutrinos! The factor $x_{\nu}\ll1$ is small for any neutral meson state. Assuming for neutrino
masses that $m_\nu \sim \sum_i m_{\nu_i} < 0.62$~eV~\cite{Goobar:2006xz}, where $m_{\nu_i}$ is the mass of one of
the neutrinos, Eq.~(\ref{eqn:Bto2nu}) yields the branching ratios of ${\cal B}_{th}(B_{s}^0 \to \nu\bar\nu) = 3.07
\times10^{-24}$, ${\cal B}_{th}(B_{d}^0 \to \nu\bar\nu) = 1.24\times10^{-25}$, and ${\cal B}_{th}(D^0 \to \nu\bar
\nu) = 1.1\times10^{-30}$ for $B_{s}$, $B_{d}$, and $D^{0}$ states, respectively. This led many authors to conclude
that ``invisible'' decays of heavy meson states are, in fact, background-free modes for searches for new light WILLPs.

\begin{center}
\begin{figure}
\center
\includegraphics[scale=1.1]{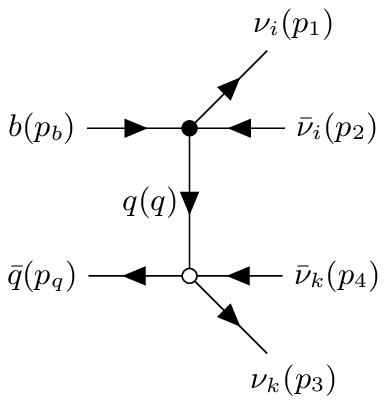}
\hspace{0.5in}
\includegraphics[scale=1.1]{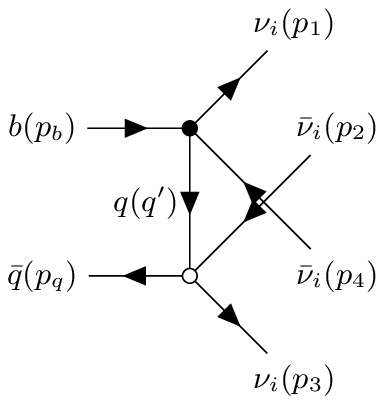}
\caption{(a) One of two diagrams for $B_{q}\rightarrow\nu_{i}\bar{\nu}_{i}\nu_{k}\bar{\nu}_{k}$ where $i,k$ refer to
different neutrino flavors. (b) Example of an additional diagram that appears in case of identical neutrinos in the
final state. Black dots represent flavor-changing, and white dots represent flavor-conserving vertices. \label{Bto4nu}}
\end{figure}
\end{center}

Here we shall point out that the $\nu \bar \nu$ final state does not constitute a good representation of invisible
width of $B_q^0 (D^0)$ mesons in the standard model. In supporting our claim we shall concentrate on the $B_q$ meson
decays, presenting the corresponding results for the $D$ and $K$ states at the end of this letter. Indeed, in the SM the
final state that is not detectable in a flavor factory setup contains an arbitrary number of neutrino pairs,
\begin{equation}\label{BtoME}
\mathcal{B}\left(B_{q}\rightarrow\slashed{E}\right)=\mathcal{B}\left(B_{q}\rightarrow\nu\bar{\nu}\right)
+\mathcal{B}\left(B_{q}\rightarrow\nu\bar{\nu}\nu\bar{\nu}\right)+...,
\end{equation}
As discussed above (see Eq.~(\ref{eqn:Bto2nu})), decay to the $\nu\bar \nu$ final state is helicity-suppressed.
The four-neutrino final state, on the other hand, does not suffer from such suppression, so it is expected to have a
considerably larger branching ratio. Naively,
\begin{equation}\label{Ratio}
\frac{\mathcal{B}\left(B_{q}\rightarrow\nu\bar{\nu}\nu\bar{\nu}\right)}{\mathcal{B}\left(B_{q}\rightarrow\nu\bar{\nu}
\right)} \sim \frac{G_F^2 M_B^4}{16 \pi^2 x_{\nu}^{2}} \gg 1.
\end{equation}
In this letter, we calculate decays of $B_{q}$, $D^{0}$, and kaons into a four neutrino final state. There are only
two diagrams that contribute to the decay amplitude when the final state has neutrinos of different flavors. Fig.\
\ref{Bto4nu}a shows one such diagram, while the other can be obtained by replacing the light-quark propagator
with a $b$-quark propagator and switching the order of vertices. We also consider the case where the final state
neutrinos are flavor identical, in which case additional diagrams appear. Fig.\ \ref{Bto4nu}b shows one such
additional diagram.
The decay amplitude $\mathcal{A}_{q}$ for $B_q \to \bar\nu \nu\bar\nu \nu$ decay can be written as
\begin{equation}\label{eqn:M1}
\mathcal{A}_{q}  = -\frac{G_{F}^{2}\alpha V_{tq}^{*}V_{tb}X\left(x_{t}\right)}{4\pi\sin^{2}\theta_{w}} \sum_{i,k} L_{\ell_{i}}^\mu L_{\ell_{k}}^\nu
\langle 0 | \bar{q}\Gamma_{\mu\nu}b | B_q \rangle
\end{equation}
where $L_{\ell_{i}}$ are the neutrino currents,
\begin{equation}\label{eqn:Ll1}
L_{\ell_{i}}^\mu \equiv\ \bar{u}^{\ell_{i}}\left(p_{i}\right)\gamma^{\mu}\left(1-\gamma^{5}\right)v^{\ell_{i}}\left(p_{i+1}\right).
\end{equation}
The effective vertex functions $\Gamma^{\mu\nu}=\sum_i \Gamma_{i}^{\mu\nu}$ are given by a combination of vertices and propagators of the light
and $b$ quarks, $i=q,b$. A particular example for the diagram Fig.~\ref{Bto4nu}a is given by
\begin{equation}\label{eqn:Gs1}
\Gamma_{q}^{\mu\nu}=
\gamma^{\nu}\left(c_{V}-c_{A}\gamma^{5}\right)\frac{\left(\ensuremath{\slashed{q}+m_{q}}\right)}{q^2-m_{q}^{2}}
\gamma^{\mu}\left(1-\gamma^{5}\right)~,
\end{equation}
where $q = p_{3}+p_{4}-p_{q}$,
$c_{V}^{b}=c_{V}^{s}=-\frac{1}{2}+\frac{2}{3}\sin^{2}\theta_{w} \equiv c_{V}$ and $c_{A}^{b}=c_{A}^{s}=-\frac{1}{2} \equiv c_{A}$. The other diagrams
with the $q$-quark propagator have a similar structure, only differing by the definition of momentum $q$, which follows from momentum conservation.
The second set of diagrams is obtained by switching the order of flavor-conserving and flavor-violating vertices.  

We adopt a simple model for calculating the transition matrix elements \cite{Aditya:2012ay},
\begin{equation}\label{eqn:MatrixElem}
\left\langle 0\right|\bar{q}\Gamma^{\mu\nu}b\left|B_{s}\right\rangle =\int_{0}^{1}dx \ {\rm Tr}\left[\Gamma^{\mu\nu}\psi_{B}\right]
\end{equation}
where $x = p_b/P_B$ is the momentum fraction of the heavy bottom quark in the $B_{q}$ meson. Here $p_b$ ($P_B$) represents the momentum
of the b-quark ($B_q$ meson). The wave function $\psi_{B}$ for the $B_q$ meson can be defined as
\begin{equation}\label{eqn:WaveFunc}
\psi_{B}=\frac{I_{c}}{\sqrt{6}}\phi_{B}\left(x\right)\gamma^{5}\left(\slashed{P}_{B}+M_{B}g_{B}\left(x\right)\right) ~,
\end{equation}
where $\phi_B(x)$ represents the quark distribution amplitude and $I_c$ is an identity in color space. We will use $g_{B}\approx1$,
which is a good approximation for a heavy meson \cite{Aditya:2012ay}. The distribution amplitude $\phi_{B}$ contains the $B_q$ meson
decay constant ($f_B$) and can be expressed as
\begin{equation}\label{eqn:phi}
\phi_{B}\left(x\right)=\frac{f_{B}}{2\sqrt{3}}\delta\left(1-x-\xi\right),
\end{equation}
which represents the fact that most of the momentum within the $B_q$ meson, $1-\xi$, is carried by a $b$-quark.
Note that $\xi$ is a small parameter, in the case of a $B_q$ meson $\xi \approx 0.1$, which allows for expansion of our results in power series in $\xi$. 
In this Letter we shall calculate the leading term in such expansion, neglecting the masses of the light quark and the neutrinos.

The kinematic region for the four-neutrino phase space depends on five independent variables.
We follow \cite{Altmannshofer:2008dz} to define two Mandelstam variables $s_{ij}$ and three
helicity angles as our independent variables. Defining $p_{ij} = p_i + p_j, q_{ij} = p_i - p_j$ for $\{i,j\} = 1-4$, we
note that $s_{ij} =p_{ij}^2 = - q_{ij}^2$, and $p_{ij}\cdot q_{ij} = 0$. The Mandelstam variables used in this Letter are $s_{12} = \left(p_{1}+p_{2}\right)^{2}$ and $s_{34}
= \left(p_{3}+p_{4}\right)^{2}$, which represent the squared invariant masses of pairs of neutrinos. Then,
$p_{12}\cdot p_{34} =(M_{B_q}^{2}-s_{12}-s_{34})/2$. The helicity angles are then defined in the
center-of-momentum frames for each Mandelstam variable. There are thus two polar angles ($\theta_Y, \theta_Z$) and one azimuthal angle ($\theta_X$),
\begin{eqnarray}\label{eqn:scalprod}
&&p_{12}\cdot q_{34} ~=~ \lambda(M^2_B,s_{12},s_{34}) \cos\theta_{Z}/2 ~,~~ \nonumber \\
&&p_{34}\cdot q_{12} ~=~ \lambda(M^2_B,s_{12},s_{34}) \cos\theta_{Y}/2 ~,~~ \nonumber \\
&&q_{12}\cdot q_{34} ~=~ (M_{B_q}^{2}-s_{12}-s_{34})\cos\theta_{Y}\cos\theta_{Z}/2 ~~~  \\
&&\hspace{2truecm} +~ \sqrt{s_{12} s_{34}} \sin\theta_{Y}\sin\theta_{Z}\cos\theta_{X} ~,~~ \nonumber \\
&&p^\mu_{12} q^\nu_{12} p^\rho_{34} q^\sigma_{34}\epsilon_{\mu\nu\rho\sigma} ~=~ - \lambda(M^2_{B_q},s_{12},
s_{34}) \sqrt{s_{12} s_{34}} ~~~ \nonumber \\
&&\hspace{3truecm} \times \  \sin\theta_{Y}\sin\theta_{Z}\sin\theta_{X}/2 ~,~~
 \nonumber
\end{eqnarray}
where $\lambda^2(x,y,z) = x^2+y^2+z^2-2xy-2yz-2xz$. In terms of these independent variables the four-body phase space with
massless final state particles takes the form,
\begin{align}\label{eqn:psA}
d \Phi =& \frac{S}{16 M_{B} \left(4\pi\right)^{6}}  d\theta_X d\cos\theta_Y d\cos\theta_Z ds_{34} ds_{12}
\end{align}
where $S=\frac{1}{j!}$ for each group of $j$ identical particles in the final state. While performing the integrals over
the four-body phase space we allow the following ranges for the helicity angles: $0 \le \theta_X \le 2\pi, 0 \le \theta_{Y,Z} \le \pi$, and the
Mandelstam variables: $0 \le s_{34} \le (M_{B_q} - \sqrt{s_{12}})^2, 0 \le s_{12} \le M_{B_q}^2$. Note that, although the integrals over the helicity angles
can be performed in any order, the allowed kinematic ranges for the Mandelstam variables were chosen in such a manner that $s_{12}$ is the final
variable to be integrated over. As expected, the final result for the decay rate does not depend on this choice of the order of integration,
which we perform numerically.

In order to find the total decay rate we consider all three flavors of neutrinos in the final state. There are six different possibilities where
the flavors of the two $\nu\bar\nu$ pairs in the final state are different, i.e. all four final state particles are distinguishable. In addition,
there are three cases where the two $\nu\bar\nu$ pairs have the same flavor, and hence there are two pairs of identical particles in the final
state. We evaluate the rate for each of the two possibilities separately and add them together with appropriate factors (factor of 6 for the former
and 3/4 for the latter) to obtain the total decay rate.

To leading order in the expansion in $\xi$ we find that $\Gamma(B_s\to\nu\bar\nu\nu\bar\nu) = (2.32 \pm 0.38) \times 10^{-27}$ GeV. Using a
$B_s$ meson lifetime of $\tau_{B_{s}}=1.509$ ps \cite{Patrignani:2016xqp}, this gives a ${\cal B}(B_s\to\nu\bar\nu\nu\bar\nu) = (5.48 \pm 0.89) \times 10^{-15}$,
which is {\it nine} orders of magnitude larger than the SM contribution to $B_{q}^0(D^0) \to \slashed{E}$ from the $\nu\bar\nu$ final state. Similarly, we find that
$\Gamma(B_d\to\nu\bar\nu\nu\bar\nu) = (6.54 \pm 1.19) \times 10^{-29}$ GeV. With $\tau_{B_{d}} = 1.520$ ps from Ref \cite{Patrignani:2016xqp},
therefore, we find ${\cal B}(B_{d}\to\nu\bar{\nu}\nu\bar{\nu}) = (1.51 \pm 0.28)\times 10^{-16}$. For the $D^{0}$ meson decay to four neutrinos, we
find that the decay rate is $(4.75 \pm 0.63)\times 10^{-39}$ GeV, and using $\tau_{D^{0}} = 410.1$ fs from Ref.\ \cite{Patrignani:2016xqp}, the branching ratio for the
four-body process is ${\cal B}(D^0\to\nu\bar{\nu}\nu\bar{\nu}) = (2.96 \pm 0.39) \times 10^{-27}$. The quoted uncertainties stem from 
the numerical calculations of phase space integrals. Even though these results
are challenging to access experimentally, they are many orders of magnitude larger than the corresponding two body decays 
$B_{d}^0(D^0) \to\nu\bar{\nu}$ \cite{Badin:2010uh}, which also contribute to $B_{q}^0(D^0) \to \slashed{E}$,
owing to powerful helicity suppression of the $B^0_{q} (D^0) \to \nu\bar{\nu}$ transitions. 

We extend our calculations to also include decays of neutral kaons to four neutrinos. We find that the corresponding decay rates for the $K_S^0$ and
the $K_L^0$ to be respectively $(4.13 \pm 0.57)\times 10^{-39}$ GeV and $(3.50 \pm 0.63)\times 10^{-39}$ GeV. Once again using the lifetimes for the
neutral kaon initial states ($\tau_{K_{S}^0}=0.8954$ ps, and $\tau_{K_{L}^0}=0.5116$ ns \cite{Patrignani:2016xqp}), we find the branching
ratios for the $K_S^0$ and $K_L^0$ decays to be $(5.62 \pm 0.78)\times 10^{-25}$ and $(2.72 \pm 0.49)\times 10^{-22}$ respectively.
\begin{center}
\begin{figure}
\center
\includegraphics[scale=1.1]{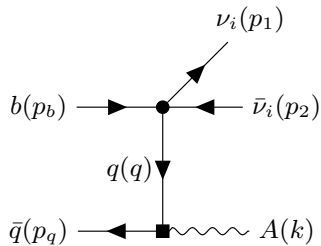}
\caption{One of the diagrams leading to $B_q^0 \to \nu\bar\nu V$ transition.
Black dot represents flavor-changing vertex (Eq.~(\ref{eqn:FCNC})), while black square represents quark couplings to $V^\prime$
(Eq.~(\ref{eqn:Couplings})).} \label{BtoNuNuV}
\end{figure}
\end{center}
In what follows we use experimental data on invisible widths of heavy mesons to constrain properties of dark photons.
The dark photons can be properly introduced in the standard model phenomenology by coupling weak isospin field $B_\mu$
to a new (dark sector) $U(1)$ vector field $V_\mu$ via kinetic mixing \cite{Holdom:1985ag},
\begin{eqnarray}\label{eqn:KinMix}
{\cal L} = &-& \frac{1}{4} W_{3\mu\nu} W^{3\mu\nu} -\frac{1}{4} B_{\mu\nu} B^{\mu\nu}
\nonumber \\
&+& \frac{\epsilon}{2} B_{\mu\nu} V^{\mu\nu} - \frac{1}{4} V_{\mu\nu} V^{\mu\nu} + \frac{m_V^2}{2} V_{\mu} V^{\mu},
\end{eqnarray}
where $\epsilon$ is the kinetic mixing parameter. The field $V_\mu$ can acquire mass via a variety of ways. After electroweak symmetry
breaking weak isospin field $B_\mu$ and the $W_{3\mu}$ combine to form the $Z$-boson $Z_\mu$ and photon $A_\mu$ fields with kinetic
mixing term ${\cal L} = (\epsilon/2) \left(\cos\theta_W F_{\mu\nu} - \sin\theta_W Z_{\mu\nu}\right) V^{\mu\nu}$.
This term can be eliminated by field redefinition \cite{Baumgart:2009tn}
\begin{eqnarray}\label{eqn:FieldRedef}
A_\mu^\prime &=& A_\mu - \epsilon \cos\theta_W V_\mu,
\nonumber \\
V_\mu^\prime &=& V_\mu + \epsilon \sin\theta_W Z_\mu.
\end{eqnarray}
The rotation of Eq.~(\ref{eqn:FieldRedef}) introduces, among other things, a direct coupling between the new ``dark photon" field
$V_\mu^\prime$ and the SM electromagnetic current,
\begin{equation}\label{eqn:Couplings}
{ \cal L} = - e \epsilon \cos\theta_W J^\mu_{em}V_\mu^\prime ,
\end{equation}
where $J^\mu_{em} = (2/3) \ \bar u \gamma^\mu u - (1/3) \ \bar d \gamma^\mu d + ...$ for up and down-type quark fields.

We can now put constraints on the parameters $\epsilon$ and $m_V$ from the experimentally constrained invisible $B_q^0(D^0)$
widths. The lowest order contribution in $\epsilon$ with invisible particles in the final state would be given by the decay
$B_q^0(D^0) \to V^\prime \bar\nu \nu$. Similarly to $B_q^0 \to \gamma \bar \nu \nu$ decay \cite{Badin:2010uh,Aliev:1996sk,Lu:1996et}
the only contributions that have no helicity suppression are the structure-dependent $V_\mu^\prime$ emissions given by diagrams of the
type pictured in Fig.~\ref{BtoNuNuV}. We use the method of calculating branching ratio for this transition as above; it has been
applied to leptonic processes with photon emission \cite{Badin:2010uh,Aditya:2012ay}. Unfortunately, current constraints on
invisible widths of heavy mesons do not yet permit placing competitive constraints on $\epsilon$. Taking the strongest
bound on $B^0_d$ invisible width and taking the massless limit of $V^\prime$ we obtain $\left|\epsilon\right| < 125$.
It is easy to show, however, that the values of $\left|\epsilon\right| \sim 3.1\times 10^{-4}$ could be probed before reaching
the ``$B_d^0 \to 4\nu$" threshold.

In conclusion, we showed that because of the helicity suppression of the two-neutrino final state, the SM contributions to invisible widths of
heavy mesons $\Gamma(B_{q}^0(D^0) \to \slashed{E})$ are completely dominated by the four-neutrino transitions $B_{q}^0(D^0) \to \nu\bar\nu\nu
\bar\nu$. This implies that invisible decays of mesons cannot be used to set constraints on neutrino masses. 
Finally, we proved that invisible decays of heavy mesons could be used to probe light DM particles \cite{Badin:2010uh}, but current
experimental bounds are insufficient to place meaningful constraints on the properties of dark photons.

This work has been supported in part by the U.S. Department of Energy under contract de-sc0007983. AAP thanks the Aspen
Center for Physics (ACP), where part of this work was performed. ACP is supported by National Science Foundation grant PHY-1607611.


\end{document}